


\font\bigbold=cmbx12


\def\pagewidth#1{\hsize= #1}
\def\pageheight#1{\vsize= #1}
\def\hcorrection#1{\advance\hoffset by #1}
\def\vcorrection#1{\advance\voffset by #1}
\newif\iftitlepage   \titlepagetrue               
\newtoks\titlepagefoot     \titlepagefoot={\hfil} 
\newtoks\otherpagesfoot    \otherpagesfoot={\hfil\tenrm\folio\hfil}
\footline={\iftitlepage\the\titlepagefoot\global\titlepagefalse
           \else\the\otherpagesfoot\fi}
\def\abstract#1{{\baselineskip=12pt\parindent=30pt\narrower\noindent #1\par
\baselineskip=16pt plus 1pt minus 1pt}}


\newcount\notenumber  \notenumber=1
\def\note#1{\unskip\footnote{\baselineskip=10pt$^{\the\notenumber}$}
{#1}\global\advance\notenumber by 1
\baselineskip=15pt plus 1pt minus 1pt}


\global\newcount\secno \global\secno=0
\global\newcount\meqno \global\meqno=1
\global\newcount\appno \global\appno=0
\newwrite\eqmac
\def\romappno{\ifcase\appno\or A\or B\or C\or D\or E\or F\or G\or H
\or I\or J\or K\or L\or M\or N\or O\or P\or Q\or R\or S\or T\or U\or
V\or W\or X\or Y\or Z\fi}
\def\eqn#1{
        \ifnum\secno>0
            \eqno(\the\secno.\the\meqno)\xdef#1{\the\secno.\the\meqno}
          \else\ifnum\appno>0
            \eqno({\rm\romappno}.\the\meqno)\xdef#1{{\rm\romappno}.\the\meqno}
          \else
            \eqno(\the\meqno)\xdef#1{\the\meqno}
          \fi
        \fi
\global\advance\meqno by1 }


\global\newcount\refno
\global\refno=1 \newwrite\reffile
\newwrite\refmac
\newlinechar=`\^^J
\def\ref#1#2{\the\refno\nref#1{#2}}
\def\nref#1#2{\xdef#1{\the\refno}
\ifnum\refno=1\immediate\openout\reffile=refs.tmp\fi
\immediate\write\reffile{
     \noexpand\item{[\noexpand#1]\ }#2\noexpand\nobreak.}
     \immediate\write\refmac{\def\noexpand#1{\the\refno}}
   \global\advance\refno by1}
\def\semi{;\hfil\noexpand\break ^^J}
\def\nl{\hfil\noexpand\break ^^J}
\def\refn#1#2{\nref#1{#2}}

\def\cmp#1#2#3{{\it Commun. Math. Phys.} {\bf {#1}} (19{#2}) #3}

\def\pl#1#2#3{{\it Phys. Lett.} {\bf {#1}B} (19{#2}) #3}
\def\np#1#2#3{{\it Nucl. Phys.} {\bf B{#1}} (19{#2}) #3}

\def\prD#1#2#3{{\it Phys. Rev.} {\bf D{#1}} (19{#2}) #3}

\def\prp#1#2#3{{\it Phys. Rep.} {\bf {#1}C} (19{#2}) #3}



\pageheight{23cm}
\pagewidth{15.6cm}
\magnification \magstep1
\baselineskip=15pt plus 1pt minus 1pt
\parskip=2pt plus 1pt minus 1pt
\tolerance=5000


\vcorrection{ 0mm}
\hcorrection{ 0mm}


\def\tr{\mathop{\rm tr}\nolimits}
\def\Tr{\mathop{\rm Tr}\nolimits}

\def\la{\langle}
\def\ra{\rangle}
\def\pa{\partial}

\def\gmn{g_{\mu\nu}}
\def\wh{\widehat}
\def\ct{{\cal T}}


{

\refn\AP
{V.I. Arnold,
\lq\lq Mathematical Methods of Classical Mechanics\rq\rq
(Springer, Berlin, 1978)}

\refn\BS
{For a review, see, P. Bouwknegt and  K. Schoutens,
\prp{223}{93}{183}}

\refn\integ
{A.M. Perelomov, \lq\lq Integrable Systems of
Classical Mechanics and Lie Algebras\rq\rq\
(Birkh\"auser Verlag, Basel, 1990);\nl
V.I. Arnold and S.P. Novikov (Eds.), \lq\lq Dynamical
Systems VII\rq\rq
(Springer, Berlin, 1994)}

\refn\Dublin
{See, for example,
L. Feh\'er, L. O'Raifeartaigh, P. Ruelle, I. Tsutsui and A. Wipf,
\prp{222}{92}{1};\nl
Feh\'er, L. O'Raifeartaigh, P. Ruelle and I. Tsutsui,
\cmp{162}{94}{399}}

\refn\Witten
{E. Witten, \cmp{92}{84}{483}}

\refn\TF
{I. Tsutsui and L. Feh\'er, \pl{294}{92}{209}}

\refn\Dirac
{P.A.M. Dirac, \lq\lq Lectures on Quantum
Mechanics\rq\rq\ (Yeshiva, New York, 1964)}

\refn\Ply
{M.S. Plyushchay, \np{362}{91}{54}; \nl
Yu.A. Kuznetsov and M.S. Plyushchay, \np{389}{93}{181}}

\refn\Br
{L. Brink, P. Di Vecchia and P. Howe,
\np{118}{77}{76}}

\refn\Bergmann
{V. Bergmann,
{\it Ann. Math.} {\bf 48} (1947) 568}

\refn\JN
{R. Jackiw and V.P. Nair, \prD{43}{91}{1933}}

\refn\Jorj
{G. Jorjadze and I. Sarishvili,
{\it Theor. Math. Phys.} {\bf 93} (1993) 1239}

\refn\HST
{Z. Hasiewicz, P. Siemion and W. Troost,
\lq\lq Chiral Quantization on a Group Manifold\rq\rq,
Leuven preprint, KUL-TF-93/41 (hep-th/9309099)}

}


{
\baselineskip=12pt
\null
\leftskip=11cm{\noindent
INS-Rep.-1042\hfill\break
DIAS-STP-94-22\hfill\break
July 1994
}
\vskip 5mm
\vfill
}

{
\baselineskip=12pt

\centerline{\bigbold Quantization of a relativistic particle}
\smallskip
\centerline{\bigbold on the $SL(2,{\bf R})$ manifold}
\smallskip
\centerline{\bigbold based on Hamiltonian reduction}

\vskip 30pt
\centerline
{
{
 G. Jorjadze
}
}
\vskip 3mm
\centerline
{\it Tbilisi Mathematical Institute}
\centerline
{\it Rukhadze 1, 380093, Tbilisi}
\centerline
{\it Georgia}
\vskip 5mm
\centerline
{
{
 L. O'Raifeartaigh
}
}
\vskip 3mm
\centerline
{\it Dublin Institute for Advanced Studies}
\centerline
{\it 10 Burlington Road, Dublin 4}
\centerline
{\it Ireland}
\vskip 5mm
\centerline
{
{
 I. Tsutsui\footnote*{E-mail: tsutsui@ins.u-tokyo.ac.jp}
}
}
\vskip 3mm
\centerline
{\it Institute for Nuclear Study}
\centerline
{\it University of Tokyo}
\centerline
{\it Midori-cho, Tanashi-shi, Tokyo 188}
\centerline
{\it Japan}
}

\vskip 50pt
\abstract{%
{\bf Abstract.}\quad
A quantum theory is constructed for the system of a relativistic
particle with mass $m$ moving freely
on the $SL(2,{\bf R})$ group manifold.  Applied to
the cotangent bundle of $SL(2,{\bf R})$,
the method of Hamiltonian reduction
allows us to split the reduced system into two coadjoint
orbits of the group.
We find that the
Hilbert space consists of states given by
the discrete series of the unitary irreducible representations of
$SL(2,{\bf R})$, and with a positive-definite, discrete
spectrum.
}

\vfill\eject


\secno=1 \meqno=1
\noindent
{\bf 1. Introduction:}\quad
In the past few years the method of Hamiltonian reduction
[\AP] has become increasingly popular and has been used,
most notably,
in the field of ${\cal W}$-algebras [\BS] and integrable models
[\integ].
The basic idea of the method is to construct a system
with certain properties out of a much simpler Hamiltonian system
with symmetry by a reduction using constraints.
For example, a large class of ${\cal W}$-algebras
can be constructed from the Kac-Moody (current) algebra [\Dublin],
whose field theoretic version is the reduction of the
(generalized) Toda theories from
the Wess-Zumino-Novikov-Witten (WZNW) models [\Witten].
Other applications
to two dimensional field theories,
including the model of
non-abelian chiral bosons, have also been reported [\TF].

In the present paper we
investigate the problem of the motion of
relativistic particles on Lie group manifolds, both
classically and quantum mechanically.
This problem is of
interest from the point of view of constraint theory [\Dirac]
because the motion of a free relativistic particle
on a manifold involves a
constraint analogous to the mass-shell condition $p^2=m^2$
in Minkowski space, and the question is how
it should be handled, especially with regard to quantization.
The problem is also of interest from the point of view of
reparametrization invariance and indeed is the
particle analogue of two dimensional conformal field theory.
The reasons for considering group manifolds in particular
are that
they are among the simplest curved manifolds and that
their analogues in two dimensional conformal field theory are
the WZNW models, where the method of Hamiltonian reduction
has been particularly useful.  Indeed, we shall see that Hamiltonian
reduction allows us to quantize the system in a rather
trivial manner, at least when the group is $SL(2, {\bf R})$.

The paper is organized as follows: We first consider general
manifolds and summarize how the Lagrangian and Hamiltonian formalism
is implemented for the reparametrization invariant theory.
Then we specialize
to manifolds corresponding to semi-simple Lie groups $G$, where
there is a left-right Noether symmetry analogous
to that in the WZNW models with conserved currents $L$ and $R$.
We shall find
that on these manifolds the constraint corresponding to the
Minkowski mass-shell condition is just
$\Tr L^2 \equiv \Tr R^2 = m^2$ --- which stipulates
our Hamiltonian reduction --- and also provide
the general solution of the reduced classical equations.
We then consider the special group
$G = SL(2, {\bf R})$ which is a three dimensional Lorentzian manifold.
This group has the property that the above
Hamiltonian reduction
leads to a split reduced system consisting of
two chiral (\lq left' and \lq right') sectors, which are both
coadjoint orbits of the group specified by the constraint.
An important
consequence is that the quantization of the system is
then reduced to finding unitary irreducible representations of
the group $SL(2,{\bf R})$.
The time-like nature of the constraint, $m^2>0$, restricts these
representations to the discrete series.
As a result,
we find that the energy levels are
positive definite and integrally spaced, while
the angular momentum takes integer values
only\note{The irreducible representations of $SL(2,{\bf R})$
were used earlier [\Ply] in constructing a quantum theory
of a relativistic particle in flat three dimensional
Minkowski space,
with curvature and torsion of a particle world trajectory.}.

\vskip 3mm
\secno=2 \meqno=1
\noindent
{\bf 2. Relativistic particle on a manifold
as a constrained system:}\quad
Let $M$ be a (pseudo-)Riemannian manifold with metric $\gmn(x)$
where $x^\mu$ is a local coordinate system on $M$.
Take the familiar action describing a relativistic
point particle of mass $m > 0$ moving freely
on the manifold $M$,
$$
I_0 = - m \int dt \sqrt{\gmn(x)\, \dot x^\mu \dot x^\nu} \,,
\eqn\rp
$$
where $t$ is a parameter along the trajectory $x^\mu(t)$ and
$\dot x^\mu := dx^\mu/dt$.
We assume that $t$ increases monotonically, say, from
$t = 0$ to $t = T$, and that paths under consideration
satisfy $\gmn(x) \dot x^\mu \dot x^\nu > 0$.
It is known that, at the classical level, one can replace (\rp)
by the quadratic action [\Br]
$$
I = - {1\over2} \int dt \Bigl[ {1\over \lambda}
     \gmn(x) \dot x^\mu \dot x^\nu
     + \lambda m^2 \Bigr] \,,
\eqn\qa
$$
with $\lambda = \lambda(t) > 0$ being a Lagrange multiplier.
Indeed, if we substitute $\lambda$
by using its equation of motion, the action $I$ reduces to
$I_0$.  Like $I_0$, the action $I$ is invariant
under reparametrizations $t \rightarrow f(t)$
with
$$
\lambda(t) \longrightarrow \lambda'(f(t))
   = \Bigl( {{df}\over{dt}} \Bigr)^{-1} \lambda(t)\,,
\qquad
x^\mu(t) \longrightarrow {x^\mu}'(f(t)) = x^\mu(t)\,,
\eqn\reparamet
$$
where we assume
$\dot f(t) > 0$ to preserve the monotonic property.

The Hamiltonian that
corresponds to the action $I$ is found to be
$$
H = - {\lambda\over 2} (g^{\mu\nu} p_\mu p_\nu - m^2),
\eqn\hammfd
$$
where $p_\mu$ is the momentum conjugate to $x^\mu$.
Since the momentum $\pi$ conjugate to $\lambda$ vanishes,
following Dirac's approach [\Dirac] to constrained systems
we must have the consistency condition $\dot \pi = \{\pi,\, H \}
\approx 0$.  This leads to
$$
\phi := g^{\mu\nu} p_\mu p_\nu - m^2 \approx 0,
\eqn\constmfd
$$
{\it i.e.}, the Hamiltonian (\hammfd) be zero.
Being first class, the constraint (\constmfd)
generates a local gauge symmetry, which is
none other than
the reparametrization of the system.  Accordingly,
the reduced phase space
is given by factorizing
the constrained surface with respect to the gauge symmetry.

\vskip 3mm
\secno=3 \meqno=1
\noindent
{\bf 3. Group manifolds:}\quad
Now we shall consider the case where $M$ is the manifold of
a semi-simple Lie group $G$, which possesses the
nondegenerate metric
$$
\gmn(x) := \Tr \Bigl( g^{-1}\pa_\mu g \, g^{-1} \pa_\nu g \Bigr)\,,
\eqn\met
$$
where $g = g(x) \in G$ is a group element.
The \lq $\Tr$' in (\met) is defined
by the matrix trace \lq $\tr$'
in some irreducible representation
multiplied by a constant $c$, so as to
provide an innerproduct $\la X\,,  Y \ra := \Tr (XY) = c \tr (XY)$
with a proper sign in the Lie algebra ${\cal G}$ of the group.
(The constant $c$ possesses a typical scale factor, which
we set to unity for brevity.)
Choosing a basis $\{ T_m \}$ in ${\cal G}$,
we have
the \lq flat' metric
in the Lie algebra,
$\eta_{mn} := \la T_m\,, T_n \ra$,
which is the metric in the
tangent space on
the group manifold\note{
As usual, $X_m := \la T_m\,, X\ra$ for $X \in {\cal G}$
and the indices are raised/lowered
as $X^m = \eta^{mn} X_n$ using the inverse $\eta^{mn}$ of the metric
$\eta_{mn}$, whence
$\la X\,, Y \ra = \eta_{mn} X^m Y^n$.  In terms of the
vielbein $e^m_{\,\,\mu} := \la T^m\,, g^{-1} \pa_\mu g \ra$
one has
$\gmn = e^m_{\,\,\mu} e^n_{\,\,\nu} \eta_{mn}$.
}.
With (\met) the action (\rp) can be written as
$$
I_0 = - m \int dt \sqrt{ \Tr (g^{-1} \dot g)^2 }\,,
\eqn\acta
$$
which is coordinate free and hence
globally well-defined over the group manifold.
The equations of motion derived from (\acta) are
$$
{d\over{dt}} \Bigl( {{g^{-1}\dot g}\over \rho} \Bigr) = 0\,,
\qquad \hbox{where} \quad
\rho :=  \sqrt{ \Tr (g^{-1} \dot g)^2 }\,.
\eqn\em
$$
Similarly, the action (\qa) admits the global form,
$$
I = - {1\over2} \int dt \Bigl[ {1\over \lambda}
     \Tr ( g^{-1} \dot g)^2  + \lambda m^2 \Bigr] \,.
\eqn\actb
$$
A salient feature of $M$ being a group manifold is that,
in addition to the reparametrization invariance,
the system acquires a chiral invariance.
In fact, both of the actions, (\acta) and (\actb),
are manifestly invariant
under the rigid left-right
transformations,
$$
g(x) \longrightarrow h g(x)\,,
\qquad
g(x) \longrightarrow g(x) \tilde h\,,
\eqn\trgl
$$
for arbitrary elements $h$, $\tilde h \in G$.

To provide a globally defined
Hamiltonian description,
let us recall the
free Hamiltonian
system that can be defined to a semi-simple Lie group $G$,
that is, the system whose
phase space ${\cal M}$ is given by the cotangent
bundle [\AP, \integ],
$$
{\cal M} =
{\rm T}^*G \simeq G \times {\cal G} = \{(g, R) \vert \,\, g \in G,\,
R \in {\cal G} \}\,,
\eqn\ps
$$
on which the symplectic 2-form is given by
$$
\omega = d\theta\,, \qquad \hbox{with} \quad
\theta = - \Tr R (g^{-1} dg)\,,
\eqn\symplectic
$$
while the Hamiltonian is
$$
H_{\rm F} = {1\over2} \Tr R^2\,.
\eqn\ham
$$
(We again set a scale constant to unity in (\ham) for simplicity.)
The non-vanishing
Poisson brackets derived from (\symplectic) are
$$
\{ R_m\,, R_n \} = f^{\quad \, l}_{mn} R_l\,, \qquad
\{ R_m\,, g_{ij}\} =  (g T_m)_{ij}\,,
\eqn\pb
$$
where $f^{\quad \,l}_{mn}$ are the structure constants
appearing in the basis:
$[T_m\,, T_n] = f^{\quad \, l}_{mn} T_l$.
Then, the Hamiltonian system we are after is furnished by
imposing the constraint (\constmfd), which now reads
$$
\phi = \Tr R^2 - m^2 \approx 0\,.
\eqn\constr
$$
Thus the (total) Hamiltonian
can be written as $H = - {\lambda\over2} \phi$
with $\lambda$ a Lagrange multiplier, which yields
the equations of motion,
$$
\dot g  \approx \{ g\,, H\} = \lambda g R\,,
\qquad
\dot R  \approx \{ R\,, H\} = 0\,.
\eqn\hm
$$
Since the constraint (\constr)
together with the first equation of (\hm)
imply
$\lambda^2 = ({\rho\over m})^2$,
the equations motion (\hm) reproduce (\em).

The conserved \lq right' current $R$
appearing in (\hm)
is in fact the Noether current
associated with the global right symmetry in (\trgl) for the action
(\acta).  Analogously,
the \lq left' current
$$
L := - g \,R \,g^{-1}\,,
\eqn\lcurrent
$$
is the conserved Noether current
associated with the left symmetry in (\trgl), which forms
the Poisson brackets,
$$
\{ L_m\,, L_n \} = f^{\quad \, l}_{mn} L_l\,, \qquad
\{ L_m\,, g_{ij}\} =  - (T_m g)_{ij}\,,
\eqn\pbl
$$
and commutes with the right current, $\{ L_m\,, R_n\} = 0$.
Both of the two currents commute with the constraint (\constr)
and are hence gauge (reparametrization) invariant.

Although unnecessary so far
in the present group manifold case,
a local coordinate system
may be useful when we wish to find
a physical interpretation for the currents.
Consider, for example,
the
normal coordinates\note{This parametrization
is available only
for a neighbourhood of the identity $g = 1$, but
this is not important for our purpose here.}
$$
g(x) = e^{x^m T_m}\,,
\eqn\ncord
$$
where $x^m$ are the \lq flat' coordinates specifying
the position of the particle.
Then, the momentum $p_m$ conjugate to $x^m$ reads
$$
p_m = - (g^{-1}\pa_m g)_n\, R^n = (\pa_m g\, g^{-1})_n\, L^n.
\eqn\momentum
$$
If we now define
the \lq vector current' $V_m$
by subtracting the two chiral currrents, we get
$$
V_m = {1\over2}(L_m - R_m) =  p_m + {\cal O}(x^2)\,,
\eqn\veccrt
$$
where ${\cal O}(x^2)$ denotes a polynomial which is
at least quadratic in $x^m$.  This shows that
in the vicinity of the origin $g = 1$ the
vector current $V_m$ reduces to $p_m$, but
since $V_m$ are conserved (while
$p_m$ are not), and since $V_m$ are  gauge invariant and survive
the reduction,
we may regard $V_m$ as the \lq momentum' (hence $V_0$ is the
\lq energy') of the particle
in the chronological gauge $x^0(t) = t$.
On the other hand,
the \lq axial vector current' $A_m$ defined by adding the
two currents becomes
$$
A_m = {1\over2}(L_m + R_m) = {1\over2} f_{mn}^{\quad \,l} x^n p_l
       + {\cal O}(x^2)\,.
\eqn\axcrt
$$
As we shall see shortly, for $G = SL(2,{\bf R})$
the current $A_m$ will be interepreted as
the generator of three dimensional
Lorentz transformations (hence
$A_0$ is the \lq angular momentum').
The orthogonality
$\la V\,,  A\ra = 0$,
which follows from (\lcurrent),
is consistent with this interpretation.

We wish to remark at this point on the general solution
for the equations of motion (\em).
Thanks to the reparametrization invariance,
the general solution can readily be found
by choosing
the invariant length for the parameter $t$ so that
$\rho = 1$.  Indeed, the equations of motion (\em)
then reduce to ${d \over {dt}} (g^{-1} \dot g) = 0$,
which can be integrated at once to be $g(t) = g(0) e^{- t R/m}$,
where $R \in {\cal G}$ is a constant satisfying (\constr).
The general solution for (\em) can
be obtained simply by returning to the generic parameter by
a reparametrization transformation $t \rightarrow f(t)$:
$$
g(t) = g(0) e^{- f(t) R/m}\,.
\eqn\solution
$$
The constant $R$ is in fact the conserved right current
determined from the initial condition, $g(0)$ and $\dot g(0)$.
(The solution can also be given in terms of
the left current as $ g(t) = e^{f(t) L/m} g(0) $.)
Thus, in the normal
coordinates (\ncord) the particle's trajectory is just
a straight line for the initial condition $g(0) = 1$.

\vskip 3mm
\secno=4 \meqno=1
\noindent
{\bf 4. Hamiltonian reduction for $G = SL(2,{\bf R})$:}\quad
We now specialize to the case $G = SL(2,{\bf R})$ which
is a three dimensional
Lorentzian manifold isomorphic to
$S^1 \times {\bf R}^2$.
We shall work with the following
basis $\{ T_m \}$ in the algebra ${\cal G} = sl(2,{\bf R})$,
$$
T_0 =
\left(
\matrix{
0 & -1 \cr
1 &  0 \cr
}
\right)\,,
\qquad
T_1 =
\left(
\matrix{
0 & 1 \cr
1 & 0 \cr
}
\right)\,,
\qquad
T_2 =
\left(
\matrix{
1 &  0 \cr
0 & -1 \cr
}
\right)\,.
\eqn\basis
$$
Choosing $c = - {1\over2}$ we find that the flat metric becomes
$$
\eta_{mn} = \la T_m\,, T_n\ra = - {1\over2} \tr (T_m T_n)
          = \hbox{diag}\,(+1, -1, -1)\,.
\eqn\flat
$$
Since the basis elements
satisfy the relation,
$$
T_m T_n =  - \eta_{mn}\cdot 1 + \epsilon_{mn}^{\quad\, l} T_l\,,
\eqn\relation
$$
with $\epsilon_{012} = +1$,
we have for $X$, $Y \in sl(2,{\bf R})$ the useful formula,
$$
XY = - \la X\,, Y\ra \cdot 1 + {1\over2} [X\, , Y] \,,
\eqn\formula
$$
and, in particular,
$X X = - \vert X \vert^2 \cdot 1 $ where
$\vert X \vert^2 := \la X\,, X \ra$.
It is then easy to show that, if we write
$X = \alpha \wh X$ with a \lq normalized' vector ({\it i.e.},
$\vert \wh X \vert^2 = \pm 1$ or 0),
we have
$$
e^X = \cases{
            \cos{\alpha}\cdot 1 + \sin {\alpha}\cdot \wh X,
                 & if $\vert \wh X \vert^2 = +1$; \cr
            \cosh{\alpha}\cdot 1 + \sinh {\alpha}\cdot \wh X,
                 & if $\vert \wh X \vert^2 = -1 $; \cr
             1 + \alpha \wh X, & if $\vert \wh X \vert^2 = 0$. \cr}
\eqn\exponent
$$
We note that the orthochronous Lorentz group
$SO^{\uparrow}(2,1)$
in three dimensions is realized by the adjoint action of $SL(2,{\bf R})$,
$$
X \longrightarrow g\, X \, g^{-1}\,, \qquad
{\rm with} \quad g \in SL(2,{\bf R})\,.
\eqn\adjoint
$$
More explicitly,
the transformations in components induced by the adjoint
action (\adjoint) read
$$
X_m \longrightarrow \Lambda_m^{\,\,\,n} X_n\,, \qquad
{\rm with} \quad \Lambda_m^{\,\,\,n} = \Tr (T_m g \,T^n g^{-1})\,,
\eqn\lorentz
$$
where
the matrices $\Lambda_m^{\,\,\,n}$
belong to the group $SO(2,1)$,
whereas the property $\Lambda_0^{\,\,0} \geq 1$
can be seen by a direct computation.
Clearly, the axial vector current
(\axcrt), which now
takes the form
$A_m =  \epsilon_{mn}^{\quad\, l} x^n p_l$,
is the generator of the Lorentz transformation (\lorentz),
and in particular $A_0$ is the angular momentum.

We now carry out the reduction of the Hamiltonian system
explicitly
by means of the constraint (\constr) in the
$SL(2,{\bf R})$ case.
The first point to be noted is that
the reduced phase space
${\cal M}_{\rm red}$ splits up into two
coadjoint orbits of the group.
To see
this,
let us first write
the variable $R \in {\cal G}$ in (\ps) used for the
phase space ${\cal M}$ as
$$
R = h^{-1} K \,h\,,
\qquad\hbox{where}\quad h \in G, \quad
K \in {\cal G},
\eqn\parar
$$
where $K$ is some fixed vector.
The parametrization (\parar)
is based on the observation
that any element in ${\cal G} = sl(2,{\bf R})$
can be reached from $K$
by an $SO^\uparrow(2,1)$ transformation
(\adjoint) with $h$, if we provide
three types of $K$, that is, time-like $\vert K \vert^2 > 0$,
space-like $\vert K \vert^2 < 0$ and null $\vert K \vert^2 = 0$.
Since one can write $K = r\wh K$ with $r > 0$ and a normalized vector
$\wh K$, one sees that
the phase space ${\cal M}$ can be parametrized by
the (redundant) set $\{g, h, r; s \}$, where
$s := \vert \wh K \vert^2 = \pm 1$,
0 indicates the type of $K$.
Substituting (\parar) back into
(\symplectic) and renaming $g h^{-1}$ as $g$, we obtain
$$
\theta = \theta_K(g) + \theta_{-K}(h^{-1})\ ,
\qquad\hbox{where}\quad
\theta_K(g) := -\Tr K(g^{-1} dg)\ .
\eqn\split
$$
If $K$ is constant but not null, then
$\theta_K$ is just the standard canonical 1-form associated with
the coadjoint orbit ${\cal O}_K$ of the group $G$
passing through $K$.  But since the constraint (\constr)
does indeed render
$K$ time-like constant with $r = m$,
we see that
the reduced phase space is given by the direct product
of the two coadjoint orbits,
$$
{\cal M}_{\rm red} \simeq {\cal O}_K \times {\cal O}_{-K}\,,
\eqn\dprod
$$
where the symplectic structure is carried over to those
on the orbits.  Accordingly,
natural variables parametrizing the reduced
phase space ${\cal M}_{\rm red}$ are
the currents on the coadjoint orbits,
$$
L = - g\, K \,g^{-1}
\qquad\hbox{and}\qquad
R = h^{-1} K \,h\,,
\eqn\coacurrent
$$
which form independently
an $sl(2,{\bf R})$ algebra under the Poisson brackets derived from
(\split).

Before going over to the quantization of the system,
we point out that for $SL(2,{\bf R})$
one can express the symplectic 2-form $\omega_K =  d \theta_K$
(or $\omega_{-K} = d \theta_{-K}$) for the coadjoint
orbit solely in terms of the chiral current $L$ (or $R$)
in (\coacurrent).
For example, in terms of the left current
the corresponding
symplectic 2-form can be written as
$$
\omega_K(g) = - {1\over{4m^2}}\epsilon^{mnl} L_m dL_n \wedge dL_l\,.
\eqn\redsym
$$
To see this, we introduce
a parameter $\delta \geq 0$ by
$\la \wh K\,, \wh L \ra : = - \cosh \delta$ with
the normalized left current $ \wh L := L/\sqrt{\vert L \vert^2}$,
and construct the three vectors,
$$
\ct_0 :=   {{\wh K - \wh L}\over{2 \cosh{(\delta/2)}}}\ ,
\qquad
\ct_1 := - {{\wh K + \wh L}\over{2 \sinh{(\delta/2)}}}\ ,
\qquad
\ct_2 := {{[\wh K\,, \wh L]}\over{2 \sinh \delta}}\ .
\eqn\uvec
$$
For fixed $\wh K$ and $\wh L$,
these vectors form a new orthonormal basis of the $sl(2,{\bf R})$ algebra,
$$
\la \ct_m\,, \ct_n \ra = \eta_{mn}
\qquad\hbox{and}\qquad
[ \ct_m\,, \ct_n ] = 2\, \epsilon_{mn}^{\quad\, l} \ct_l\,.
\eqn\onbasis
$$
With this basis we consider the Euler angle representation
of $SL(2,{\bf R})$ elements,
$$
g = g(\alpha,
\beta, \gamma) =
     e^{\alpha \ct_2}\, e^{\beta \ct_0}\, e^{\gamma \ct_2}\,.
\eqn\euler
$$
Note that among the three parameters is a bounded one
$0 \leq \beta <2\pi$,
which is the parameter in the cyclic direction $S^1$
of the group manifold
$SL(2,{\bf R})$ (see (\exponent)).
Observe also that the Lorentz transformation on the vector
$\wh K$ by the adjoint
action of $g(\alpha) = e^{\alpha \ct_2}$ is a \lq rotation'
in the plane spanned by $\wh K$ and $\wh L$,
$$
\wh K \longrightarrow g(\alpha)\, \wh K \, g^{-1}(\alpha)
  =  {{\sinh{(2\alpha +
\delta)}}\over{\sinh\delta}} \wh K
  +  {{\sinh{2\alpha}}\over{\sinh\delta}} \wh L\ .
\eqn\rotation
$$
One then finds that for $\alpha = - \delta/2$ the
vector $\wh K$ is rotated to $-\wh L$, and for
$\alpha = -\delta/4$ it is rotated
halfway to $-L$, {\it i.e.,}
it directs to $\ct_0$.
But since the parametrization (\euler)
consists of two rotations of the
type (\rotation) with $g(\alpha)$ and $g(\gamma)$,
interrupted by
the rotation with $e^{\beta \ct_0}$, the
parameters fulfilling the relation $L = - g K g^{-1}$ in
(\coacurrent) are found to be
$$
\alpha = \gamma = -
{\delta\over 4}\,, \qquad \beta = \hbox{arbitrary}\,.
\eqn\answ
$$
(The appearance of the free
parameter $\beta$ is expected from
the counting of degrees of freedom ---
$SL(2,{\bf R})$ is three dimensional
while its coadjoint orbit is two dimensional for
$m \ne 0$.)
If we now express the canonical 1-form $\theta_K$
in (\split) using (\euler) and (\answ), we get
$$
\theta_K(g)  = - m d\beta +
  {{ \la [\wh K\,, L]\,, d L \ra}\over
   {4(m - \la \wh K\,,  L \ra)}}\,.
\eqn\canonical
$$
Choosing, {\it e.g.,}
$\wh K = - T_0$ we find that the corresponding
symplectic 2-form
$\omega_K$ is just the one given in
(\redsym).  Note that from (\coacurrent)
this choice implies
$$
L_0 > 0 \qquad \hbox{and} \qquad R_0 < 0,
\eqn\signs
$$
that is, the left current lies in the
coadjoint orbit given by an upper
hyperboloid in the algebra $sl(2,{\bf R})$ whereas
the right current lies in the coadjoint orbit given by a lower one.

\vskip 3mm
\secno=5 \meqno=1
\noindent
{\bf 5. Quantization:}\quad
We are now going to discuss the quantization of the system.
However, having seen that the reduced phase space consists of
the two coadjoint orbits (\dprod),
the problem actually
reduces to the quantization of the system of
coadjoint orbits.
In other words, the quantization amounts
to finding unitary, irreducible
representations of the algebra $sl(2,{\bf R})$ formed by
the chiral currents
on the coadjoint orbits,
${\cal O}_K$ and ${\cal O}_{-K}$.
On account of the constraint (\constr)
which requires the Casimir
$ q = {1\over4} \Tr L^2 $ to be
positive constant ${{m^2}\over4}$,
the irreducible
representations
[\Bergmann] (see also, [\JN]) relevant for our purpose
are the discrete series
$D^\pm_j$ with $2j = 3, 4, \ldots\,$,
for which $q = j (j - 1) > 0$.
Further, the conditions (\signs) require that
the representations for the left
sector should be given by $D^+_j$ while those for the
right sector are $D^-_j$.
A simple realization for these representations
can be provided by the Holstein-Primakoff method, in which
one uses creation/annihilation operators
$
[a \,, a^\dagger] = 1
$
as a basic building block.  For instance,
for the left sector we have [\Jorj],
$$
\eqalign{
L_- &:= L_1 + iL_2 = 2\sqrt{a^\dagger a + 2j}\cdot a\,, \cr
L_+ &:= L_1 - iL_2 = 2 a^\dagger
\cdot \sqrt{a^\dagger a + 2j}\,, \cr
L_0 &:= 2(a^\dagger a + j).
}
\eqn\curpar
$$
It is straightforward to check that
the left current given in (\curpar) satisfies
the constraint (\constr) as well as
the (quantum) commutation relations,
$$
[ L_m\,, L_n ] = 2 i \, \epsilon^{\quad \, l}_{mn} L_l\,.
\eqn\com
$$
In the familiar Fock space
consisting of the states
$\vert n_{\rm L} \ra$ for $ n_{\rm L} = 0, \, 1, \, 2, \ldots$ with
$$
a \vert n_{\rm L} \ra = \sqrt n_{\rm L} \,\vert n_{\rm L}-1 \ra\,,
\qquad
a^\dagger \vert n_{\rm L} \ra = \sqrt{n_{\rm L}+1} \,\vert n_{\rm L}+1 \ra\,,
\eqn\caop
$$
we find
$$
\eqalign{
L_- \,\vert n_{\rm L} \ra
   &= 2\sqrt{(n_{\rm L}-1+2j)n_{\rm L}}\, \vert n_{\rm L}-1 \ra\,, \cr
L_+ \,\vert n_{\rm L} \ra
   &= 2\sqrt{(n_{\rm L}+2j)(n_{\rm L}+1)}\, \vert n_{\rm L}+1 \ra\,, \cr
L_0 \,\vert n_{\rm L} \ra
   &= 2(n_{\rm L}+j)\, \vert n_{\rm L} \ra\,. \cr
}
\eqn\dr
$$
Analogously, one can construct representations
for the right sector using
another pair of creation/annihilation operators
for the right current.
Actually, this is equivalent to
the formal replacement
$\{ L_+, L_-, L_0 \} \rightarrow \{ - R_-, -R_+, -R_0 \}$
in the above construction, which leads to
the Fock space consisting of
$\vert n_{\rm R} \ra$ for
$ n_{\rm R} = 0, \, 1, \, 2, \ldots$, for which
$$
\eqalign{
R_- \,\vert n_{\rm R} \ra &=
 -2\sqrt{(n_{\rm R}+2j)(n_{\rm R}+1)}\, \vert n_{\rm R}+1 \ra\,, \cr
R_+ \,\vert n_{\rm R} \ra &=
 -2\sqrt{(n_{\rm R}-1+2j)n_{\rm R}}\, \vert n_{\rm R}-1 \ra\,, \cr
R_0 \,\vert n_{\rm R} \ra &=
 -2(n_{\rm R}+j)\, \vert n_{\rm R} \ra\,. \cr
}
\eqn\rdr
$$

The full Hilbert space
is spanned by the states given by the
direct product of the
two representations,
$D^+_j$ and $D^-_j$,
sharing the same value for the Casimir.  The states
are thus labeled by two integers,
$\vert n_{\rm L}\,, n_{\rm R} \ra = \vert n_{\rm L} \ra \otimes
\vert n_{\rm R} \ra$, on which the energy $V_0$
in (\veccrt) and the angular mementum $A_0$
in (\axcrt) act as
$$
\eqalign{
V_0\, \vert n_{\rm L}\,, n_{\rm R} \ra &=
 (n_{\rm L} + n_{\rm R} + 2j)\, \vert n_{\rm L}\,, n_{\rm R} \ra\,, \cr
A_0\, \vert n_{\rm L}\,, n_{\rm R} \ra &=
 (n_{\rm L} - n_{\rm R})\, \vert n_{\rm L}\,, n_{\rm R} \ra\,. \cr
}
\eqn\eigen
$$
The above result shows that the energy levels are
positive definite and spaced integrally ---
which is in fact expected because of our identification
of $x^0 \in S^1$ being \lq time' ---
while the angular
momentum takes integer values only.
The allowed mass of the particle
at the quantum level is
$$
m = 2 \sqrt{j(j-1)}\,, \qquad \hbox{with}
       \quad 2j = 3,\, 4,\, \ldots.
\eqn\mass
$$

As we have seen in this paper, the basic ingredient underlying
the simplicity of the quantization is the chiral split of the
reduced system, that is, the split into two coadjoint orbits.
In this respect, it is worth mentioning that
essentially the same split was discussed recently
(for compact groups) in [\HST] for the system
of the cotangent bundle.
This suggests that
the Hamiltonian reduction and the
subsequent quantization considered for $SL(2,{\bf R})$
may be generalized to any higher rank group $G$
with the simplicity intact, by
specifying all the Casimir elements of the
group in the form of constraints.
Whether this yields a physically interesting model
or not is however unclear except for
$G = SL(2, {\bf R})$.

\vskip 3mm
\noindent
{\bf Acknowledgement:}
One of us (G.J.) is indebted to the Dublin Institute
for Advanced Studies and the Max-Planck-Institut
f\"ur Physik, M\"unchen,
where part of this work has been done, for kind hospitality.

  \vfill\eject\immediate\closeout\reffile
  \centerline{{\bf References}}\bigskip\frenchspacing%
  \input refs.tmp\vfill\eject\nonfrenchspacing

\bye